\input epsf
\documentstyle[iopconf1]{article}
\newcommand{\ApJ}{{\em Astrophys. J.} }
\newcommand{\AandA}{{\em Astron. Astrophys.} }
\newcommand{\AandAS}{{\em Astron. Astrophys. Suppl.} }
\newcommand{\CommA}{{\em Comments Astrophys.} }
\newcommand{\EPJ}{{\em Europhys. J.} }
\newcommand{\GRG}{{\em Gen. Rel. Grav.} }
\newcommand{\HEAD}{{\em AAS High Energy Astrophys. Div.} }
\newcommand{\MN}{{\em Mon. Not. R. astr. Soc.} }
\newcommand{\NewA}{{\em New Astron.} }
\newcommand{\NewAR}{{\em New Astron. Rev.} }
\newcommand{\PASJ}{{\em Proc. Astron. Soc. Japan\/} }
\newcommand{\PASP}{{\em Proc. Astron. Soc. Pacific\/} }
\def\spose#1{\hbox to 0pt{#1\hss}}
\def\simlt{\mathrel{\spose{\lower 3pt\hbox{$\mathchar"218$}}
     \raise 2.0pt\hbox{$\mathchar"13C$}}}
\def\simgt{\mathrel{\spose{\lower 3pt\hbox{$\mathchar"218$}}
     \raise 2.0pt\hbox{$\mathchar"13E$}}}

\begin{document}

\title{New physics from the Cosmic Microwave Background}
\author{Douglas Scott\footnote{E-mail: dscott@astro.ubc.ca}}

\affil{Department of Physics \& Astronomy,
University of British Columbia,
Vancouver, B.C. V6T 1Z1~~Canada}

\beginabstract
I review the present status of the Cosmic Microwave Background,
with some emphasis on the current and future implications for
particle physics.

\endabstract

\section{Background}

Figure~1 gives an overview of information on background radiation
in the Universe.  The reason to plot $\nu I_\nu$ is so that it is possible
to read off the relative contributions to total energy density.
What can be seen is that the Cosmic Microwave Background (CMB) is by far
the dominant background.  The CMB corresponds to an energy density of
$0.260\,{\rm eV}\,{\rm cm}^{-3}$, or a number density of $410\,{\rm cm}^{-3}$,
corresponding to about 2 billion photons per baryon in the Universe
today.  On the figure, the next biggest background
-- almost two orders of magnitude
down in energy  contribution -- is in the far-IR/sub-mm part of the
spectrum, and believed to come from distant, dusty, star-forming galaxies.
A little below that is the near-IR/optical background, coming from the sum
of the emission of all the stars in all the galaxies we can observe.  Then
much lower are the X-ray and $\gamma$-ray backgrounds, which come predominantly
from active galactic nuclei.

\begin{figure}
\begin{center}
\epsfxsize=12.5cm \epsfbox{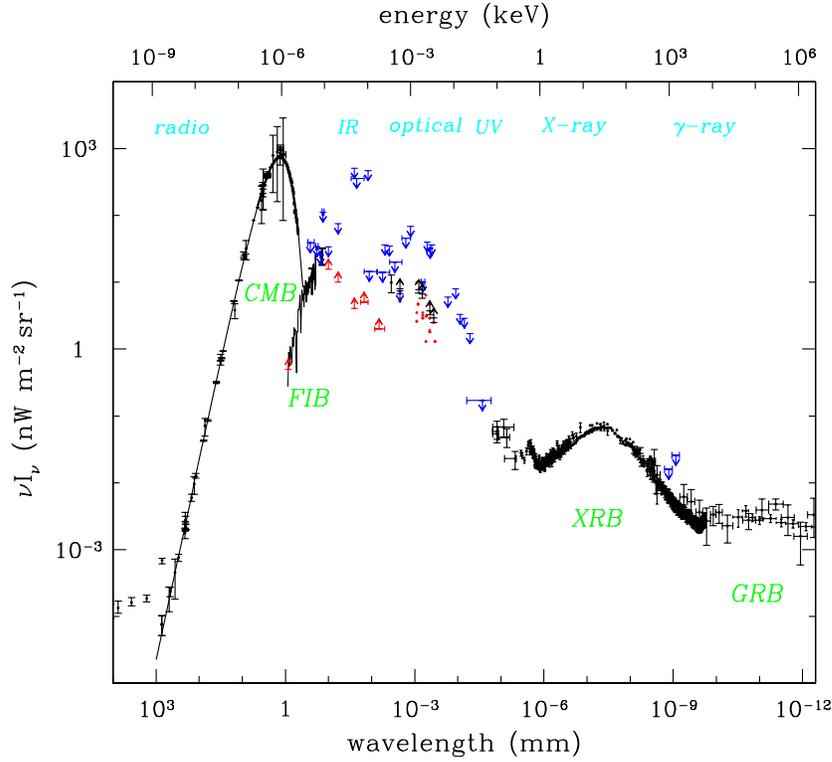}
\end{center}
\footnotesize\rm
\caption{A compilation of recent constraints on extragalactic diffuse
background radiation.  In terms of total energy the CMB dominates, with the
Far-Infrared and Optical Backgrounds about a factor of 100 lower.
These data are based upon the older compilation of primarily from Ressell \&
Turner~1990 \cite{ResTur}, supplemented with more recent data from:
Smoot~1997 \cite{Smoot97} for the CMB; Lagache \etal 1999 \cite{Lag99} and
Hauser \etal 1998 \cite{Hauser} for the FIB;
Leinert \etal 1998 \cite{Leinert} for a near-IR to near-UV compilation;
Dwek \& Arendt~1998 \cite{DweAre} for the near-IR;
Pozzetti \etal 1998 \cite{Poz} for the optical;
Miyaji \etal 1998 \cite{Miy} and Gendreau \etal 1995 \cite{Gen} for the X-ray;
and Sreekumar \etal 1998 \cite{EGRET},
Kappadath \etal 1999 \cite{COMPTEL} and Watanabe \cite{SMM}
for the $\gamma$-ray.
In the colour version lower limits are shown in
red and upper limits in blue.}
\end{figure}

Apart from the CMB, there is no evidence for background emission arising from
anything other than known sources of radiation: stars, gas and dust
within galaxies.  In other words, there is no reason to believe that decaying
particles, for example, distributed throughout the Universe, are
contributing much to the background, and hence these sorts of data can place
constraints on exotic particles (e.g.~\cite{ResTur}).
The CMB is different, however.  Its
spectral shape is spectacularly well-fit by a blackbody
\cite{FIRASCOBRA,Fixsen,Smoot97}, over more than 4 decades in frequency.
The current best estimate of the CMB temperature is
\begin{equation}
T_0 = 2.725 \pm 0.001\ {\rm Kelvin}
\end{equation}
\cite{Mather99}.  The fact that the CMB is such a good blackbody is one of the
pillars of the standard Big Bang model.  The argument is
that, since we can't even make such a good blackbody in the lab, the CMB
needs to have originated from something in extraordinarily good thermal
equilibrium.  The only known source is the entire Universe, during an
earlier epoch when it was very much hotter and denser.  Together with
the Hubble law for distant galaxies, this leads to a robust model
in which the Universe used to be hot and dense, and has been cooling and
expanding since then.

Since we know that the Universe consists mainly
of hydrogen, we can calculate (see \cite{SSS} for a recent update)
that the Universe was ionized
when it was hotter than about $4{,}000\,$K (a lower temperature than you
might have first guessed, because of the high photon-to-baryon ratio).  Since
the radiation redshifts just like $T\propto(1+z)$, then that means the
Universe recombined\footnote{This process is always called
{\em re}combination, even although in the cosmological context the atoms
begin by being {\em un}combined, and hence the process is really
combination.  For particle physicists who feel the corners of their mouths
lifting here, let me point out that this is not too dissimilar to the talk
of symmetry restoration in the early Universe, which only makes literal
sense if time runs backwards.} at $z\simeq1500$.  This was the time when
the radiation last interacted with the matter (through electron scattering),
and most CMB photons have been travelling freely since then.  In models
with typical parameters, this epoch corresponds to a time of around
$300{,}000$ years.

The spectrum is thermalised (by double Compton and Bremsstrahlung) for
times earlier than about $1\,$year, corresponding to $z\,{\sim}\,10^7$.
Hence particle decays, or other energy emitting processes, occurring over
the redshift range $10^3\,{<}\,z\,{<}\,10^7$,
could leave an observable signature on the
CMB spectrum (see e.g.~\cite{SmoSco}).

Although all measurements are currently only upper limits, there is
some prospect of detections of spectral distortions from planned spectral
experiments \cite{HalSco}.  For example,
at low frequencies it seems feasible to detect Bremsstrahlung emission from
the reionized gas in the inter-galactic medium at moderate redshifts
\cite{DIMES}.
However, progress in constraining other realistic physical effects
will require considerably greater improvements in experimental sensitivity.

As well as distortions to the spectrum, the CMB also contains cosmological
information in the variations in temperature across the sky
\cite{Reviews}.
After the detection of CMB anisotropy by the {\sl COBE\/} satellite in
1992 \cite{Smo92} attention has been focussed almost exclusively on these
anisotropies.  This is partly because it became clear that detection
was easily within reach of state-of-the-art detectors, but also because
theoretical calculation showed that precise measurements of the anisotropy
power spectrum would provide detailed information about fundamental
cosmological parameters \cite{Boltz}.

\begin{figure}
\begin{center}
\epsfxsize=11.5cm \epsfbox{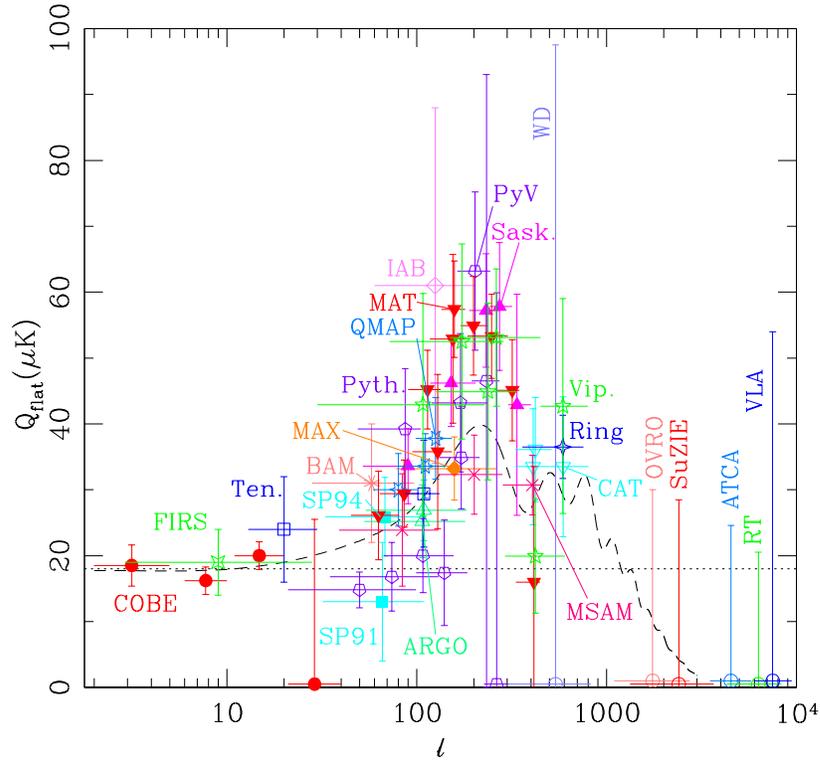}
\end{center}
\footnotesize\rm
\caption{Most of the experiments published to date.  See Smoot \& Scott
(1997) \cite{SmoSco} for full references, supplemented with more recent
results from: OVRO Ring \cite{Ring}, QMAP \cite{QMAP}, MAT TOCO \cite{MAT},
CAT \cite{CAT}, Python V \cite{PyV} and Viper \cite{Viper}.  The error
bars (these are $1\sigma$ except for the upper limits which are 95\%)
have generally been symmetrised for clarity, and calibration
uncertainties are included in most cases. The horizontal bars represent
the widths of the experimental window functions.  The dotted line is the
flat power spectrum which best fits the {\sl COBE\/} data alone.  The dashed
curve is the prediction from the vanilla-flavoured
standard Cold Dark Matter model.}
\end{figure}

Since {\sl COBE\/}
there have been around 20 separate experiments which have detected
temperature fluctuations which are most likely to be primordial.  These
are summarized in Figure~2.
Here the $x$-axis is the spherical harmonic multipole, $\ell$.  The temperature
fluctuation field on the sky can be decomposed into an orthogonal set of
modes:
\begin{equation}
{\Delta T\over T}\left(\theta,\phi\right)
 = \sum_{\ell,m} a_{\ell m} Y_{\ell m}(\theta,\phi).
\end{equation}
Since there is no preferred direction on the sky
(e.g.~\cite{BunSco}) the individual $m$s are irrelevant, and so the
important information is contained in the power spectrum
\begin{equation}
C_\ell \equiv \left\langle \left| a_{\ell m} \right|^2 \right\rangle.
\end{equation}
Indeed if the perturbations are Gaussian, then this contains {\it all\/} the
information.
The conventional amplitude of the quadrupole is given as
\begin{equation}
{Q^2\over T_0^2} \equiv {\sum_m \left| a_{\ell m} \right|^2\over4\pi}
= {5C_2\over 4\pi}.
\end{equation}
A `flat' spectrum means one in which
$\ell(\ell+1)C_\ell={\rm constant}$, and we can therefore define that
constant in terms of the expectation value for the 
equivalent quadrupole $Q_{\rm flat}$ -- which
is what is plotted as the $y$-axis in Figure~2.

Each experiment quotes one
(or in the best cases several) measures of power over a range of multipoles,
and these can be quoted as `band powers' or equivalent amplitudes of a flat
power spectrum through some `window function'.  The horizontal bars on the
points are an indication of the widths of these window functions.

\section{What do we know?}

There are a number of things to note from Figure~2 \cite{LawScoWhi}:
\begin{itemize}
\item The plot has become very crowded!
\item The overall detection of
anisotropy is at the ${\simeq}\,40\sigma$ level.
\item 
A flat power spectrum (horizontal dotted line) is a bad fit, at
about the $15\sigma$ level.
\item
There is clear evidence for a peak at $\ell\sim200$.
\end{itemize}

What do we learn from this?  First of all, it looks like our basic
paradigm -- to describe the large scale properties of the Universe,
and the formation of structure within it -- are in good shape.  The
prediction from the `straw man' model, standard Cold
Dark Matter (sCDM) is shown by the dashed line in Figure~2.
The point is that this model, which contains parameters which
are all fixed at very round numbers, has the right general character.  And
it is easy to find models which fit the data much better, by tuning some
of those parameters.

What we already know can be split (artificially) into 3 areas:
astrophysics, cosmology and particle physics, each of which I will now
discuss in turn.  For more details refer to the review article by
Lawrence et al.~(1999) \cite{LawScoWhi}.

\subsection{Astrophysics}

Before the {\sl COBE\/} detection, there had been about 25 years of quoted
upper limits to CMB anisotropy.  There was much talk in the literature about
how new paradigms would be required if {\sl COBE\/} returned yet stronger
upper limits.  But in fact the detection was just at the right level for
gravity alone to have grown the structure from amplitudes of
${\sim}\,10^{-5}$ at $z\,{\sim}\,1000$ to the non-linear structures we see
today.  This is easy to arrange for models in which the Universe is
dominated by non-baryonic dark matter, and with adiabatic perturbations.
Thus, probably the most important thing to come out of the {\sl COBE\/}
anisotropy measurements (apart from the general good news that we are
on the right track!) is the realization that
\begin{center}
{\it Gravitational instability in a dark matter dominated universe
grew today's structure}
\end{center}

This `fact', arising from the CMB,
has added to the Big Bang paradigm, so that the picture is
of a hot expanding Universe, which at early times contained
small amplitude density perturbations.  The obvious next question is where
those perturbations came from -- an issue we shall return to in a minute.

\begin{figure}
\begin{center}
\epsfxsize=11.5cm \epsfbox{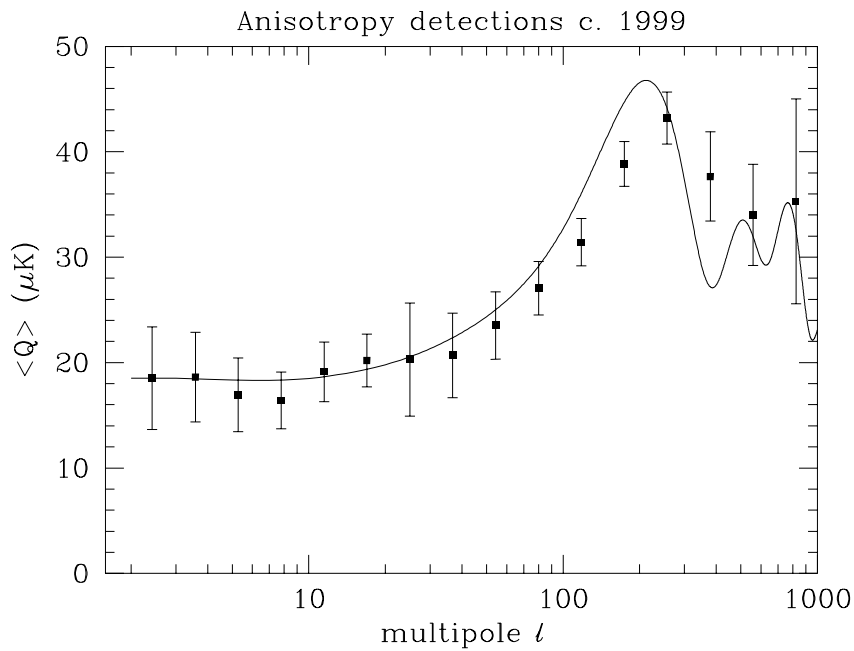}
\end{center}
\footnotesize\rm
\caption{The result of binning the data in the previous figure.
More precisely, what was done was to split the multipoles into
16 bins between $\ell=2$ and $\ell=1000$, and to weight each experiment
by the fraction of the window in each bin.  The precise height of the
peak depends to some extent on the choice of bins, on details of the
window functions used, and on the weights given to individual experiments.
The points here are not un-correlated, but provide a reasonable visual
impression of the current data -- more sophisticated treatments
(e.g.~\cite{BonJafKno} give similar results).
The solid line is a $\Lambda$-dominated CDM model, with parameters which
are consistent with most current cosmological constraints.
}
\end{figure}

Meanwhile, there are a few other things that the current suite of CMB
measurements tells us.  First of all, it is pleasing that the
approximate scale of the peak (apparent in the binned plot, Figure~3)
is just where it is theoretically predicted
in simple models.  This acoustic peak corresponds to the length scale which a
sound wave can travel at the time of recombination, projected onto
the sky, and was contained in papers at least as early as Doroshkevich,
Sunyaev \& Zel'dovich (1978) \cite{DSZ}.  The position in angle is also
a good test for the geometry of the Universe, since it comes from the
projection of a fixed physical scale onto the sky.

We know (from the lack of complete absorption
shortwards of the Lyman edge in distant quasars) that most of the material
in the nearby Universe is ionized, out to redshifts $z\,{>}\,5$.
Whether the Universe reionized at $z\,{\sim}\,10$ or $z\,{\sim}\,1000$ is,
however, not obvious.  But, very early reionization would lead to the erasing
of the small-scale CMB anisotropies, which patently has {\it not\/}
occurred.  Hence we can infer that
\begin{center}
{\it The Universe remained neutral until $z\,{\simlt}\,50$}
\end{center}
(see e.g.~\cite{GriBarLid}).

\subsection{Cosmology}
Let us imagine for the sake of this section, that `cosmology' is synonymous
with the search for the values of a number of parameters which describe
the properties of the Universe.

Figure~3 appears to show that
\begin{center}
{\it The CMB power spectrum peaks at
$150\,{\simlt}\,\ell\,{\simlt}\,350$,}
\end{center}
avoiding any detailed statistical arguments here, and just sticking to
round numbers (and remembering that there are tight upper limits at
higher values of $\ell$, so that the power spectrum really does have
to come down again).  Since the standard
CDM model has the main peak at $\ell\,{\simeq}\,220$, and it is pushed to 
smaller scales in open models,
then it is hard for $\Omega_{\rm tot}$ to be less than,
say, 0.3.  Rigorous studies (e.g.~\cite{DodKno}) arrive at similar
conclusions.

The height of the peak is somewhat higher than predicted for sCDM, but
entirely consistent with several variants\footnote{Rather than vanilla CDM
you can have a slightly different flavour, or some chocolate sprinkles, or
maybe a cherry on top.}.  Currently popular models with a cosmological
constant tend to provide perfectly good fits to the CMB (in addition to
large-scale clustering of galaxies and the supernovae results
\cite{SN}).  The curve
plotted in Figure~3, shows one such flat model with $\Omega_\Lambda=0.6$
and a Hubble constant of $70\,{\rm km}\,{\rm s}^{-1}{\rm Mpc}^{-1}$
\cite{Freedman}.

Since the height of the first peak depends on a combination of 
parameters, then exactly what quantities are constrained
depends on the parameter space being searched, as well as on the choice
of additional constraints.  Currently it is possible to constrain
the matter density $\Omega_{\rm M}$ to ${\sim}\,\pm0.1$ from the peak height,
but that depends sensitively on the assumptions used.  All this is
expected to change as better data come in.

The basic thing to take away here is that models with adiabatic-type
(i.e.~where you perturb the matter and radiation at the same time in order
to keep the entropy fixed) perturbations have the right kind of character.
On the other hand isocurvature-type models (where the matter and radiation
get equal and opposite perturbations, so that the local curvature is
unperturbed) tend to look poor -- generically they have a `shoulder'
rather than a first peak, and then the highest peak is at much smaller
scale (see e.g.~\cite{HuSpeWhi}).
While there are some loop-holes, it seems difficult to get
isocurvature models to fit the current data.

\subsection{Particle physics}

Let me briefly discuss some particle physics implications.  For a more
complete discussion see the recent review by Kamionkowski
\& Kosowsky \cite{KamKos}.

The best-fitting models for the CMB, in conjunction with other cosmological
constraints, seem to imply that there may be a positive cosmological
constant, or other form of matter which behaves in a similar way.  Since
`extraordinary claims require extraordinary evidence' I think it is
premature to say that the energy density of the vacuum has now been
measured.  But since $\Omega_\Lambda\,{\simeq}\,0.6$ really is the best fit at
the moment, it is worth exploring this more fully.  The implications for
particle physics models are obviously profound.

The CMB has little to say about the neutrino, unless its mass is high enough
that free-streaming of the dark matter particles is important (which is
only the case for $m_\nu\,{\simgt}\,{\rm few\ eV}$, see \cite{HuEisTeg} for
discussion).
But the CMB certainly requires
(again much more strongly when other constraints
are taken into account) that most of the matter content is in some cold dark,
non-baryonic form -- and some new particle is the favoured candidate.
Whether it is the axion, the lightest supersymmetric particle, or something
else, remains to be seen.

There are many other constraints on particle physics which are beginning to
be discussed.  The basic idea is that the Universe couldn't have been too
crazy at $z\,{\sim}\,1000$, otherwise the microwave sky would appear very
different.  Already there have been limits placed on:
strong primordial magnetic fields; large domains of anti-matter; large
lepton asymmetry; particle decays; and early phase transitions.  At the
moment the limits are not too severe on things that anyone believed in the
first place, but this will certainly change as the data improve.

I'm not sure how much this belongs in the particle physics section, but there
are various models in which the large-scale structure of the Universe is
non-trivial, either in terms of global rotation or topology.  If these
are too extreme they lead to detectable patterns on the CMB sky
\cite{rotate}.
In the simplest models with strange topology, the scale has to be so
close to the Hubble scale as to be hardly worth considering
\cite{T3}, although in models with hyperbolic geometry things are much
less clear \cite{Topen}.  But of course, open models are
not currently in vogue.  In any case the conclusion is that
\begin{center}
{\it The large-scale structure of spacetime appears to be simple.}
\end{center}

\subsubsection{Inflation.}

Most people working in the field which is sometimes referred to as
`CMB phenomenology' are currently struggling with the same question, in one
form or another (e.g.~\cite{Proof}):
how does one confront the concept of Inflation with the
concept conventionally known as Proof?

It seems clear that we are now in a position to say something beyond the Big
Bang paradigm.  The CMB led us to accept that the Universe used to be
hotter and denser, and more recently to the conclusion that structure built
up through gravitational instability.  Now it appears that we are learning
something further, something about the origin of the perturbations themselves.
But just exactly {\it what\/} that next step is, and how to phrase it, is
altogether less clear.  For lack of anything better, let me phrase my own
current belief, which I challenge anyone to disagree with:
\begin{center}
{\it Something like Inflation is something like proven}
\end{center}

Of course the interpretation of this statement depends on the exact definition
of the two crucial words.  I'm sure that I don't know what I mean by
`proof'.  But by `inflation' I mean some mechanism which gave rise to a roughly
scale-invariant spectrum of adiabatic perturbations, over a wide range of
scales, including those which are apparently acausal.  The only causal
way we know of to do this is to have the scale factor accelerate
(${\ddot a}>0$) at some time in the early history of the Universe
\cite{addot}.  And we
can argue about whether something that achieves the same end result is
just isomorphic to inflation, even if interpretted differently.  `Inflation'
does not necessarily carry with it the extra baggage of an inflaton
potential etc. -- although hopefully the connection with particle physics
would follow later.

It used to be that discussions of inflation focussed on the number of
e-foldings required to solve horizon, flatness, entropy and monopole
problems.  However, at the present time the paramount concern is making those
darned density perturbations.  And inflation gives you a mechanism to do
that, for free!  It appears that we are learning that the Universe has
inflation-like `initial conditions'.  Time will tell whether that means
that the Universe was once dominated by some vacuum energy density, and
whether we can learn details about particle physics at ultra-high energies.
The promise of the CMB is that it provides a way of probing density
perturbations while they were still in the linear regime (i.e.~simple).
Thus we may be able to learn details of how the perturbations were generated
which {\it may\/} lead to direct information about physics at energies
at the GUT scale, or even the Planck scale.

\subsubsection{Defects.}

Since I had many discussions at this meeting about alternatives to inflation,
let me dwell a little on that subject here.
The only real competitor to inflation has been any one of various
field ordering mechanisms or topological defect models.  Generically these
give larger CMB anisotropies (from the so-called integrated Sachs-Wolfe
effect) for the same density perturbation amplitudes\footnote{This
is basically because of their similarity to isocurvature models; adiabatic
(i.e.~inflationary) models, on the other hand, give the correct value to a
factor of 2, without really breaking sweat.}.  The power spectra of
galaxy perturbations, or even the underlying dark matter fluctuations, are
notoriously complicated to calculate -- nevertheless there seems little
evidence that the observed power spectrum can be easily reproduced in these
sorts of models.  Moreover, there now seems to be some consensus in the view
that generic defect models produce at most one (broad) peak
in the CMB power spectrum \cite{Defects}
and in a place which tends to give a poor fit to current data.

The status of defects vs. the Universe can be summarized in the following
three points.
Defect models tend to give
\begin{itemize}
\item the wrong matter power spectrum;
\item the wrong CMB power spectrum;
\item the wrong normalization of matter relative to CMB.
\end{itemize}
But apart from that, these models seem to work fine!

There is certainly a strong motivation for working on such models simply
from the point of view that they are {\it cool}.  Some of the required
mathematics is interesting in its own right and some of the numerical
calculations are challenging.  It would be neat if the Universe was full of
a network of cosmic strings, containing within them trapped regions of
GUT-scale physics, and thrashing around at near the speed of light.
To put it another way, who wouldn't sometimes rather be Captain Jean-Luc
Picard?  But ultimately it doesn't matter what sort of Universe
we would like to live in\footnote{My own sense of humour makes models
with say $\Omega_{\rm tot}=1.1$ sound pretty appealing!},
we are stuck with this particular one, and we are learning a great deal
about its properties.  Details of the structure within our Universe seem
relatively easy to fit with inflationary-type models, and considerably
harder with defect-type models.

Which is not to say that defects are not important in other branches of
physics -- or even perhaps for other purposes in the early Universe --
but at this point they seem to hold little promise as a method of
forming structure.

\subsubsection{Other paradigms?}

There is of course an argument in favour of investigating other possibilities,
at least until such time as inflation has been more directly tested.
While the current suite of defect models do not look very promising, there
is always the possibility of a more attractive model lying around corner.

New ideas from particle physics also have the potential for providing
different mechanisms for structure formation.  Exactly what will come out
of string theory, large extra dimensions and broken Lorentz invariance
remains to be seen.  It will be interesting to see how generic the basic
inflationary predictions are, and whether new twists on high energy
physics carry with them new testable predictions.

\section{Future experiments}

More information can be found in 
recent reviews (e.g.~\cite{Law,HalSco}).  Here I give a very brief summary.

The next generation of CMB balloon experiments are expected to return data
of much higher quality (and quantity).  The new results from BOOMERanG 
\cite{BOOM} (as
well as MSAM, MAXIMA and others) are eagerly awaited.  BOOMERanG `98 was the
first long-duration balloon flight, and by all accounts was staggeringly
successful.  Three immediate questions are expected to be addressed by this
new data-set: do the currently favoured $\Lambda$-dominated cosmologies
continue to be a good fit; what is the precise location of the first peak;
and is there any evidence for other peaks.  This last point is perhaps
the most important.  Detection of oscillations in the power spectrum, with
tight constraints on the peak spacings, will be a very firm test of the
inflationary paradigm \cite{HuWhi}.

The adiabatic, apparently acausal perturbations, generated during inflation,
give a series of peaks in the ratio $1:2:3:\cdots$ in $\ell$-space.  On the
other hand, causal, isocurvature perturbations naturally give rise to peaks
in the ratio $1:3:5:\cdots$.  So detection of a second peak at roughly half
the angular scale of the first, will be a very large step towards `proving
inflation'.  Failure to observe this will, of course, be even more exciting,
since it will demand an entirely new paradigm.

In the short term there are also at least three new interferometer
projects (e.g.~\cite{DASI}): DASI at the South Pole, CBI in Chile and the
VSA in Tenerife.  All of these are nearing completion and the improvement
in the available data is expected to increase even more when they return data
within the next year or two.

Another direction being pursued from the ground is CMB polarization --
see \cite{PolExpt} for experimental details and \cite{PolTheor} for a
theory primer.  The CMB sky is naturally polarized
at the few percent level (the result of the quadrupole term in Compton
scattering together with a slightly anisotropic radiation field at
$z\,{\sim}\,1000$).  Measuring the ${\sim}\,\mu$K signals will be very
challenging, but can provide information beyond that contained in the
temperature anisotropies alone.  Since polarization is such a strong
prediction, it better be there, otherwise our whole picture has to change!
Furthermore, we can more definitively separate any
gravity wave contribution in the CMB (if it is measurable\cite{ZibScoWhi}),
thus limiting the energy scale of inflation.
Large-angle polarization can also constrain
the reionization epoch, and details of the polarization power spectrum
are a direct probe of physics around the time of last scattering.  This is
all in addition to the fact that polarization simply gives extra information
to better constrain parameters (and to break degeneracies between some
combinations of parameters).

\section{MAP and Planck}

Two satellite missions are currently planned to study the CMB from space,
where the whole sky can be imaged, far from the complicating effects of
the atmosphere.  The NASA Microwave Anisotropy Probe ({\sl MAP\/}
\cite{MAP})
is due for launch in November 2000.  It will travel to the Earth-Sun outer
Lagrange point, L2, where it will map the sky at 5 frequencies between
22 and $90\,$GHz, reaching to $\ell\,{\sim}\,800$ in the power spectrum.
The careful control of systematics possible with an
extended space mission means that {\sl MAP\/} should represent a very large
improvement over the data available from the Earth-based experiments.

The ESA mission
{\sl Planck\/} can be thought of as the third generation CMB satellite,
mapping at 9 separate frequencies between 30 and $850\,$GHz, with both
radiometer and bolometer technologies, and measuring the $C_\ell$s
to beyond $\ell$ of 2000.  Thus {\sl Planck\/} is expected
to measure essentially {\it all\/} of the primordial CMB power
spectrum (see Figure~4), and cover all the frequencies required to
measure and remove the foreground signals.  The {\sl Planck\/} data set
should enable cosmological parameters to be constrained with exquisite
precision -- or, to put it another way, the power spectrum should be
measured at a level of several million $\sigma$.
In addition {\sl Planck\/} will measure the polarization
(and cross-correlation with temperature) power spectra, providing even
more information.

\begin{figure}
\begin{center}
\epsfxsize=12cm \epsfbox{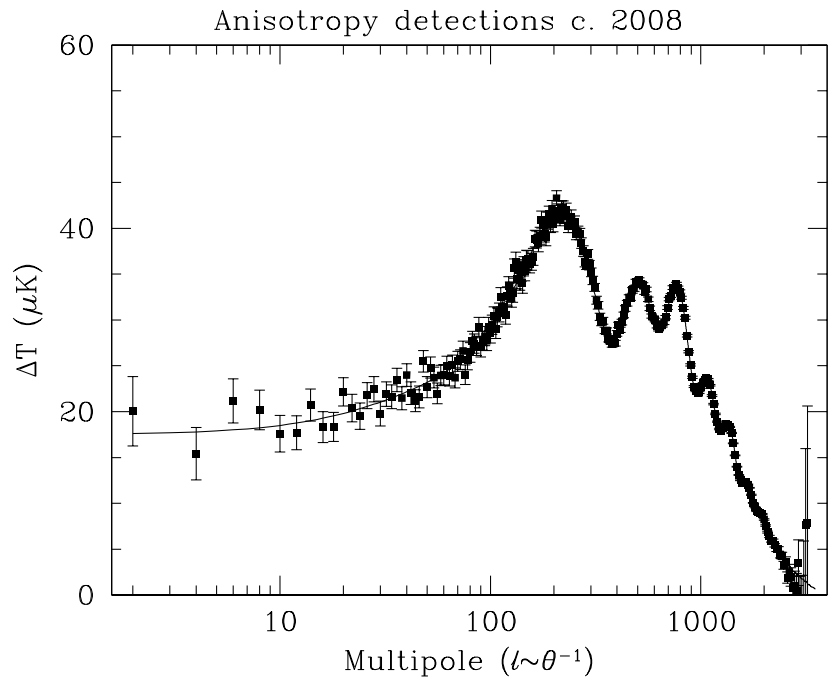}
\end{center}
\footnotesize\rm
\caption{
An estimate for how well the power spectrum might be measured by
{\sl Planck\/}.  This is a realization of a CDM power spectrum, assuming
the {\sl Planck} instrumental sensitivity over two thirds of the sky.
{\sl Planck\/} should supply us with essentially cosmic-variance limited
information on all the angular scales relevant to primary anisotropies, over
the full range of relevant frequencies.
}
\end{figure}

In terms of particle physics, there will be constraints on {\it anything\/}
which could potentially affect the anisotropies.  This is just like Big Bang
Nucleosynthesis constraining strange things occurring at ${\sim}\,$MeV
energies, or ${\sim}\,$minute timescales.
There have already been many studies (too many to list in detail)
estimating how well various things could be limited by {\sl Planck\/} data.
These include:
variation of fundamental constants; decaying particles;
$\Lambda$, Quintessence, rolling scalar fields, Dark Energy, etc. equation
of state; alternative gravity models; parity violation; extra relativistic
degrees of freedom; and just about anything else you can think of which isn't
already ruled out.

Certainly cosmological parameters will be constrained.  And definitely
some messy astrophysical details will be uncovered (in the foregrounds, as
well as through some weak processing effects occurring between
$z\,{=}\,0$ and 1000).  And whatever the basic paradigm, there will
surely be some clues to fundamental physics lurking in there, since the
CMB anisotropies provide the cleanest information about the initial
conditions and the largest scale properties of the Universe.
What is therefore clear is that
\begin{center}
{\it We will learn a great deal about cosmology, astrophysics and particle
physics from {\sl MAP\/} and {\sl Planck\/}}
\end{center}

\section{Conclusions}
The main points are highlighted in italics throughout.  We are
beginning to learn the answers to some fundamental questions,
using information contained in CMB
anisotropy data.  With future experiments, and the {\sl MAP\/} and
{\sl Planck\/} satellites in particular, we should expect to learn vastly
more in the coming years about astrophysics, cosmology and particle physics.

\section*{Acknowledgements}

I would like to thank my collaborators, and in particular members of the
{\sl Planck\/} team.  Since this is a conference proceedings report, I
have shamelessly concentrated on my own work -- see the original papers
for more comprehensive references.
I am also grateful to the members of CITA for their hospitality during the
writing of this article.

\end{document}